\documentclass[12pt,emtex]{article}%{amsart} %
\usepackage{amsfonts}
\usepackage{amssymb}
\usepackage{amscd}
\newcommand{\BEQ}{\begin{equation}}
\newcommand{\EEQ}{\end{equation}}
\newtheorem{Th}{Theorem}

\newtheorem{Prop}{Proposition}
\newtheorem{Cor}{Corollary}
\newtheorem{Conj}{Conjecture}
\newtheorem{Ex}{Example}
\newtheorem{Def}{Definition}
\newtheorem{Rem}{Remark}
\def\nn{\nonumber}
\def\one#1{#1^{\raise5pt\hbox{$\scriptstyle\!\!\!\!1$}}\,{}}
\def\two#1{#1^{\raise5pt\hbox{$\scriptstyle\!\!\!\!2$}}\,{}}

\def\bea{\begin{eqnarray}}
\def\eea{\end{eqnarray}}

\def\C{{\mathbb{ C}}}
\def\CC{{\mathbb{ C}}}

\def\l{\lambda}

\def\g{\mathfrak{gl}_n}

\def\p{\partial_z}

\begin{document}
\begin{titlepage}
\hfill ITEP-TH-10/06 \vskip 2.5cm
\centerline{\Large KZ equation, G-opers, quantum Drinfeld-Sokolov reduction}
\smallskip
\centerline{\Large and quantum Cayley-Hamilton identity  }

\vskip 1.5cm \centerline{A. Chervov \footnote{E-mail:
chervov@itep.ru} , D. Talalaev \footnote{E-mail:
talalaev@itep.ru} }

\centerline{\sf Institute for Theoretical and Experimental Physics
\footnote{ITEP, 25 B. Cheremushkinskaya, Moscow, 117259, Russia.}}
\vskip 2.0cm
\centerline{\large \bf Abstract}
\vskip 1cm
The Lax operator of the Gaudin type models is a
{\em 1-form} on the classical level.
In virtue of the quantization scheme
proposed in [Talalaev04] (hep-th/0404153) it is natural to treat the quantum
Lax operator as a {\em connection}; this connection is a particular case of the
Knizhnik-Zamolodchikov
connection [ChervovTalalaev06] (hep-th/0604128).
In this paper we find a gauge transformation
which produces the "second normal form" or the "Drinfeld-Sokolov"
form.
Moreover the differential operator naturally corresponding to this
form is given precisely by the quantum
characteristic polynomial [Talalaev04] of the Lax operator (this
operator is called the $G$-oper or Baxter equation).
This observation allows to relate solutions of the KZ and Baxter equations in an obvious way,
and to prove that the immanent KZ-equations has only meromorphic solutions.
As a corollary we obtain the quantum Cayley-Hamilton identity for the
Gaudin-type Lax operators
(including the general $\g[t]$ case).
The presented  construction sheds a new light on a geometric Langlands
correspondence. We also discuss the relation with the Harish-Chandra
homomorphism.

\vskip 1.0cm
\end{titlepage}
%\tableofcontents
\section{Introduction and main results}
\subsection{Simple fact from linear algebra}
Let $L$ be a generic matrix over a field, then it can be conjugated to
the "second normal form", i.e. there exists a matrix $C$, such that
\bea
C~L~C^{-1}=
\left(
\begin{array}{cccccc}
0 & 1 & 0 & 0 &... &0 \\
0 & 0 & 1 & 0 &... &0 \\
... & ... & ... & ...&... &... \\
0 & 0 & 0 & ... & 0 &1 \\
H_n & H_{n-1} & H_{n-2} & ... & H_2 &H_1
\end{array}\right)
\eea
where $H_i$ are the coefficients of the characteristic polynomial:
$$det( L(z)-\l)=(-1)^n(\l^n-\sum_i H_{n-i} \l^i) $$
i.e. $H_1= Tr (L)$, $H_n=(-1)^{n-1}det (L)$.
\\
To prove this statement one just takes the matrix $C$
to be the matrix of the basis change to the basis consisting of vectors:
$v, Lv, L^2v,..., L^{n-1} v$, where $v$ is a generic vector.
{\Ex Let $L$ be a numerical $2\times2$ matrix
\bea
L=
\left(
\begin{array}{cc}
a & b \\
c & d
\end{array}
\right)\nn
\eea
with $c\ne 0.$ This matrix can be transformed to the second normal form by the matrix
\bea
C=
\left(
\begin{array}{cc}
0 & 1 \\
c & d
\end{array}
\right)\nn
\eea
i.e.
\bea
CLC^{-1}=\left(
\begin{array}{cc}
0 & 1 \\
-det(L) & tr(L)
\end{array}
\right)\nn
\eea
}
\\
For the purposes of the present paper it is instructive to
rewrite the formula above in the following more general way:
\bea
\label{conj-class}
CL=\left(
\begin{array}{cc}
0 & 1 \\
-det(L) & tr(L)
\end{array}
\right)C
\Leftrightarrow
C(L-\l)=\left(\left(
\begin{array}{cc}
0 & 1 \\
-det(L) & tr(L)
\end{array}
\right)-\l\right)C
\eea

\newpage

\subsection{The quantization scheme}
Here we recall the quantization scheme for the class of integrable systems
with the rational Lax operator on the example of the Gaudin model.
\paragraph{Lax operator}
The Lax operator of this model is a rational function with distinct poles:
\bea
\label{main-part-Lax-Gaud}
L(z)=\sum_{i=1...N}
\frac{\Phi_{i}}{z-z_i}\nn,
\eea
 $\Phi_i\in Mat_n\otimes \bigoplus\g \subset Mat_n\otimes
U(\g)^{\otimes N}$ is
 defined by the formula:
\bea \label{Phi-Gaudin}
\Phi_i=\sum_{kl}E_{kl}\otimes e_{kl}^{(i)}
\eea
where $E_{kl}$ form the  standard basis
in $Mat_n$ and $e_{kl}^{(i)}$ form a basis in the $i$-th copy of $\g.$

{\Rem ~}
All our considerations identically holds true
for any other Lax operators satisfying the linear $R$-matrix
commutation relation
\ref{rel1}, in particular for the standard Lax operators for $\g[t]$ given
by the formula \ref{Lax-n}.

\paragraph{Quantum characteristic polynomial}
The quantum commutative family is constructed with the help of the quantum
characteristic polynomial
\bea \label{qchar}
"det"(L(z)-\partial_z)=Tr A_n (L_1(z)-\p)\ldots
(L_n(z)-\p)=\sum_{k=0}^nC_n^k(-1)^{n-k}QI_k(z)\p^{n-k}
\eea
{\bf Theorem \cite{T04}} {\it The coefficients $QI_k(z)$ commute
\bea
[QI_k(z),QI_m(u)]=0\nn
\eea
and quantize the classical Gaudin hamiltonians.}
\\~\\
In these formulas $L_i(z)$ is an
element of ${Mat_n}^{\otimes n}\otimes U(\g)^{\otimes N}$
obtained via the inclusion
$Mat_n\hookrightarrow
{Mat_n}^{\otimes n}$ as the $i$-th component. The element $A_n$ is the
normalized
antisymmetrization operator in ${\C^n}^{\otimes n}.$
{\Rem By the same formula in \cite{CT06-1} it was constructed a commutative subalgebra in
$U(\g)[t]/{t^N},~U(\g)[t]$ and the center of the universal enveloping affine
algebra on the critical level .}

\newpage

\subsection{KZ-equation, G-opers and Baxter equation}
The brief relation scheme is the following:
\begin{itemize}
%ACRED total
\item {\em Langlands correspondence. G-opers.}\\
The scalar differential operator given by
\bea
\chi("det"(L(z)-\partial_z))=\sum_{k=0}^nC_n^k(-1)^{n-k}\chi(QI_k(z))\p^{n-k}
\eea
is called $G$-oper in the theory of the geometric Langlands correspondence,
this defines a connection on the punctured disc (Galois side of the correspondence).
Here $\chi$ is a character of the commutative
subalgebra generated by $QI_k(z)$.
It is related to the character of the center of $U_{crit}({\widehat { \g}})$
and hence to the representation of $U_{crit}({\widehat { \g}} )$
(automorphic side of  the correspondence).
\item {\em Baxter equation. Baxter's $Q$-operator.}\\
Baxter equation and its solution (the Baxter's $Q$-operator)
are given by the formula:
\bea
\pi("det"(L(z)-\partial_z)) Q(z)=\sum_{k=0}^nC_n^k(-1)^{n-k}\pi(QI_k(z))\p^{n-k} Q(z)=0.
\eea
Here $(\pi,H)$ is a representation of the algebra $U(\g)^{\otimes N}$ in a Hilbert space $H$.
$U(\g)^{\otimes N}$ is the algebra of quantum observables of the Gaudin
model,\\
$H=V_1\otimes ... \otimes V_N$ is its Hilbert space.
$Q(z)$ is an $End(H)$-valued function.
\item {\em Knizhnik-Zamolodchikov equation.}\\
The standard KZ-equation \cite{KZ} for the particular value of the level
 is given by the formula:
\bea
\pi(L(z)-\p)\Psi(z)=
\left(\sum_{i=1...N}
\frac{ \sum_{kl}E_{kl}\otimes \pi_i(e_{kl}^{(i)})}
{z-z_i} -\p\right)\Psi(z)=0.
\eea
Here $(\pi, V_1\otimes ... \otimes V_N)$ is a representation
of $U(\g)^{\otimes N}$ and $\Psi(z)$ is a
$\CC^n\otimes V_1\otimes ...\otimes V_N$-valued function.
 In \cite{CT04} it
was shown that having a solution for the universal KZ equation
$(L(z)-\p)\Psi(z)=0$ one obtains solutions for the universal $G$-oper just
taking an arbitrary component $\Psi(z)_i$ of the vector $\Psi(z)$
($\Psi(z)_i$ is a $V_1\otimes ...\otimes V_N$-valued function.)
\end{itemize}
For more details we refer to our previous work \cite{CT06-1}.

\subsection{Main results}
\paragraph{Conjugation of the quantum Lax operator to the Drinfeld-Sokolov form}
~\\
Let $L(z)\in Mat_n \otimes U(\g)^{\otimes N} \otimes Fun(z) $ be the quantum Lax operator
of the Gaudin model, here $Fun(z)$ is an appropriate space of functions on the formal parameter
$z.$
Let us denote by $L^{[i]}(z)$ the {\em quantum powers} of the Lax operator
defined by the following formulas:
\bea
\label{quant-power}
L^{[0]}&=&Id\nn\\
L^{[i]}&=& L^{[i-1]}L+\p L^{[i-1]}\nn
\eea
{\Th
\label{main}
The element $C(z)$ of $Mat_n \otimes U(\g)^{\otimes N} \otimes Fun(z)$
given by
\bea
\label{C}
C(z)=
\left(
\begin{array}{c}
v \\
v L \\
v L^{[2]}\\
\ldots\\
v L^{[n-1]}
\end{array}\right)
\eea
where $v\in \CC^n$
provides the following gauge transformation
\bea
\label{hyp}
C(z)(L(z)-\p)=
\left(\left(
\begin{array}{cccccc}
0 & 1 & 0 & 0 &... &0 \\
0 & 0 & 1 & 0 &... &0 \\
... & ... & ... & ...&... &... \\
0 & 0 & 0 & ... & 0 &1 \\
QH_n & QH_{n-1} & QH_{n-2} & ... & QH_2 &QH_1
\end{array}\right)-\p\right)C(z)
\eea
where
\bea
"det"(L(z)-\p)=Tr A_n (L_1(z)-\p)\ldots
(L_n(z)-\p)=(-1)^n(\p^n-\sum_i QH_{n-i} \p^i)
\eea
}
\\
One denotes the connection on the right hand side  of the
Drinfeld-Sokolov type.
{\Cor The quantum powers of the Lax operator satisfy the quantum Cayley-Hamilton
identity
\bea
\label{CH}
L^{[n]}(z)=\sum_{i=1}^n  QH_i(z) L^{[n-i]}(z)
\eea}
\\~\\
{\bf Proof~}
Considering the last line of the equation (\ref{hyp}) one obtains
\bea
vL^{[n-1]}(z)(L(z)-\p)=\sum_{i=1}^n  v QH_i(z) L^{[n-i]}(z)-\p
vL^{[n-1]}(z)\nn
\eea
which is equivalent to the corollary $\square$
\\~\\
As another corollary we obtain the simple way to produce solutions of the universal $G$-oper from
solutions for the KZ equation by a linear transformation. This method is
in agreement with the one considered in \cite{CT04}.
\\~\\
{\bf Historic remarks~} The Cayley-Hamilton
identity above is the first (to the best
of our knowledge) case of such identity
where the Lax operator with the spectral parameter
is used.
The unexpected appearance of the corrections
to the powers of $L(z)$ differs it from the standard ones.
For the Lax operator without spectral parameter
the identities were discussed in \cite{HC} and
references therein (see also \cite{Molev02} section 4.2 page 37).
In  \cite{Kir} (see section 2.4 "Discussion") the
Cayley-Hamilton identity for $\g$ was treated (this paper was inspiring for us).
Let us also mention that  in \cite{GelfandBig} (section 8.6 page 96)
the generalization of the Hamilton-Cayley identity
of somewhat different nature was found for matrices with coefficients in arbitrary
noncommutative algebra.

~\\
{\bf Acknowledgements~}
The work of the both authors
has been partially supported by the RFBR grant 04-01-00702,
and the grant of Support for the Scientific Schools 8004.2006.2.
The draft idea of the theorem \ref{main} was formulated
during the stay of one of the authors (D.T.) at LPTHE in autumn 2005. D.T.
would like to thank the French Government and the direction of FRIP
(F\'ed\'eration de Recherches Interactions Fondamentales)
 for organization of this visit
and especially O. Babelon for the valuable contribution to understanding of the
problem.
The work of one of the authors (AC) has been partially supported by the INTAS
grant YSF-04-83-3396,
the part of it was done during the visit to SISSA under the INTAS project,
the author is deeply indebted to SISSA and especially to G. Falqui
for providing warm hospitality, excellent working conditions and stimulating  discussions.

\section{Main section}
\subsection{Comparison with the general  KZ$\leftrightarrow$G-oper correspondence}
In \cite{CT04} it was observed that there is a simple connection between
solutions of the KZ equation
\bea
(L(z)-\p)S(z)=0\nn
\eea
for $S(z)$ - a function which takes values in $\C^n\otimes V$ where $V$
is a representation of $U(\g)^{\otimes N};$
and of the equation produced from the quantum
characteristic polynomial
\bea
\label{UGO}
"det"(L(z)-\p)\Psi(z)=0
\eea
for $\Psi(z)$ - a function with values in $V.$ One has to take as $\Psi(z)$ the
projection to the antisymmetric part of the expression
$U(z)=v_1\otimes\ldots\otimes v_{n-1}\otimes S(z)$ where $v_i$ are arbitrary
vectors in $\C^n.$ With special choice of vectors $v_i$ one shows that all
components of $S(z)$ over $\C^n$ solve the equation (\ref{UGO}).
{\Rem The role of the equation (\ref{UGO}) is very important in the problem
of solving the quantum model and in the Langlands correspondence, namely by
restricting this equation to the common eigen-vector for the Gaudin
hamiltonians one obtains the so-called $G$-oper; the condition that this
differential equation does not have monodromy is equivalent to the Bethe
ansatz conditions for the eigenvalues of quantum hamiltonians.}
\\
~
\\
The formula (\ref{hyp}) of the main theorem also provides such a relation,
namely: if $S(z)$ solves the KZ-equation
\bea
(L(z)-\p)S(z)=0\nn
\eea
then the first vector component of $C(z)S(z)$ solves the Baxter-type (or G-oper)
 equation (\ref{UGO}).
Taking $C(z)$ in the form \ref{C} with the vector $v=(0,...,1,...,0)$
we see that the first component of $C(z)S(z)$ is just
$S(z)_i.$

{\Cor The KZ equation
\bea
(\pi (\p-L(z)))S(z)=0
\eea
 has only rational solutions
(here $\pi$ is a finite-dimensional representation of $U(\g)^{\otimes N}$).
}
\\
~
\\
{\bf Proof~} It was conjectured in \cite{CT04} (on the basis of \cite{Frenkel95} and
the ideas of Baxter, Gaudin, Sklyanin)
that the Baxter-type equation $\pi("det"(\p - L(z)))\Psi(z)=0$
has only rational solutions. The conjecture was proved in \cite{MV2}
(see theorem 4.1 page 12).
On the other hand we know from the result of \cite{CT04} discussed above that any component $S(z)_i$
of the vector $S(z)$ solves Baxter equation, hence $S(z)$
is the vector-valued rational function as well.
$\square$

\subsection{Conjugation}
Let us firstly give some examles of quantum powers of the Lax operator
(\ref{quant-power}):
\bea
L^{[1]}(z)&=&L(z)\nn\\
L^{[2]}(z)&=&L^2(z)+L'(z)\nn
\eea
Hence for the $n=2$ case the matrix $C$ can be taken just in the classical
form
\bea
C(z)=\left(
\begin{array}{cc}
0 & 1\\
c(z) & d(z)
\end{array}\right)
\eea
where the quantum Lax operator is
\bea
L(z)=\left(
\begin{array}{cc}
a(z) & b(z)\\
c(z) & d(z)
\end{array}\right)
\eea
\\~\\
{\bf Proof of theorem \ref{main}~}
\\
For the general case let us consider a solution $\Psi(z)$ for the KZ equation
\bea
L(z)\Psi(z)=\p \Psi(z)\nn
\eea
Let $\Phi(z)=C(z)\Psi(z)$ where $C(z)$ is given by the formula (\ref{C}).
Then
\bea
\Phi_1(z)&=&<v,\Psi(z)>\nn\\
\Phi_2(z)&=&<vL(z),\Psi(z)>=<v,\p\Psi(z)>\nn\\
\ldots\nn\\
\Phi_k(z)&=&<v(L^{[k-1]} L(z)+\p L^{[k-1]}),\Psi(z)>\nn\\
&=&
<vL^{[k-1]},\p \Psi(z)>+<v\p L^{[k-1]},\Psi(z)>=\p\Phi_{k-1}(z)\nn
\eea
Hence such $C(z)$ transforms the KZ connection to a Drinfeld-Sokolov type
connection
\bea
\left(
\begin{array}{cccccc}
0 & 1 & 0 & 0 &... &0 \\
0 & 0 & 1 & 0 &... &0 \\
... & ... & ... & ...&... &... \\
0 & 0 & 0 & ... & 0 &1 \\
\vartheta_n(z) & \vartheta_{n-1}(z) & \vartheta_{n-2}(z) & ... & \vartheta_2(z)
&\vartheta_1(z)
\end{array}\right)-\p
\nn
\eea
To prove that this is exactly the connection on the right-hand
side of (\ref{hyp}) one has to apply the fact due to \cite{CT04}, that each component of the KZ
solution $\Psi$ satisfy the $G$-oper. On the other hand by construction we obtain that
$<v,\Psi(z)>$ satisfy the differential equation
\bea
\label{diff}
(\p^n-\sum_{i=1}^n\vartheta_i(z)\p^{n-i})<v,\Psi(z)>=0
\eea
Due to generality of the choice of the
KZ solution $\Psi(z)$ and the vector $v\in \CC^n$ we conclude
that the differential operator (\ref{diff}) and the $G$-oper coincide up to
multiplication by a function. They coincide because their leading terms
do $\blacksquare$

\section{Factorization of QCP}
\subsection{Miura form}
{\Cor~ Let the QCP be represented in a factorized form
\bea
\label{factor-QCP}
"det"(\p - L(z))=Tr A_n(\p - L_1(z))...(\p - L_n(z))=(\p-\chi_n(z))...(\p-\chi_1(z))
\eea
 then
there exists $C^d(z)$ belonging to some algebraic extension of the quantum
algebra such that:
\bea
C^d(z)(\p - L(z))=\left(\p -
\left(
\begin{array}{cccccc}
\chi_1(z) & 1 & 0 & 0 &... &0 \\
0 & \chi_2(z) & 1 & 0 &... &0 \\
... & ... & ... & ...&... &... \\
0 & 0 & 0 & ... & \chi_{n-1}(z)&1 \\
0 & 0 & 0 & ... & 0 &\chi_{n}(z)
\end{array}\right)
\right) C^d(z)
\eea
Let us denote the connection on the right hand side by $\p-L_{M}.$
}
\\
We provide a proof in the appendix.

{\Rem  Similar factorizations of Baxter-type equations often occur in
mathematical physics literature, for the general Gaudin and XXX models
the explicit factorization of scalar $G$-opers related to our constructions
were given recently by E. Frenkel, E. Mukhin, V.Tarasov,
and A. Varchenko.}

{\Rem ~\label{RemFacFreed} In fact the differential operator $D=\sum_i H_i(z) \p^i$
can be represented in a factorized  form
$D=(\p-\chi_n(z))...(\p-\chi_1(z))$
non-uniquely.
Let $\Psi_1,\ldots,\Psi_n$ be a basis of solutions of the equation $D\Psi=0$,
let us pick out $\chi_1,\ldots,\chi_n$ recursively such that
$$(\p-\chi_i)\ldots(\p-\chi_1)\Psi_i=0$$
For example $\chi_1=\Psi_1'/\Psi_1.$
Hence different factorizations are
parametrized by the $n$-dimensional flag variety.}

{\Rem  The matrix $C(z)$ is invertible only in the field
of fractions of $U(\g)^{\otimes N}$, this means that $\pi (\p - L(z))$
and $\pi(\p- L_{DS}(z)) $ are not fully gauge equivalent,
where $\pi$ is a representation of $U(\g)^{\otimes N}$.
One can produce a solution of  $\pi( "det"(\p - L(z)))$ from any solution
of $\pi (\p - L(z))$, but this correspondence is not necessarily a
bijection.}

{\Rem
All the results of this paper can be generalized to semisimple Lie algebras
and quantum groups following the ideas discussed in \cite{CT06-1}.
For example for Lax operators related to $\g[t]$
the same theorems should hold (the Lax operator can be considered
in any representation, not only fundamental),
for Yangian: $e^{\p} - T(z)$ can be conjugated to DS-form with
the coefficients given by $"det"(e^{\p} - T(z))$, for $U_q(\hat g)$
the same is valid for $q^{\p} - L^+(q^z)$.}

\subsection{Harish-Chandra map}

In fact the factorization
formula \ref{factor-QCP} for a particular case of the QCP
provides an explicit realization of the Harish-Chandra homomorphism.
\\
Let us consider the representation $\pi,V_\l$ of $\g$
with the highest weight $\l=(\l_1,...,\l_n)$,
(i.e. that there exists a vector $|0>\in V_\l$  such that
$\pi ( e_{ii})|0>=\l_i|0>$ and $\pi (e_{ij})|0>=0$ for
$i<j$).

{\Conj Consider the $1$-spin Lax operator $L(z)=\Phi/z$ where
$\Phi\in Mat_n\otimes U(\g), ~ \Phi=\sum E_{kl}\otimes e_{kl}$
\bea
\label{spin1-factor}
\pi ( "det"(\p-L(z))) |0> ~ = ~ (\p-\l_1/z)(\p-\l_2/z)...(\p-\l_n/z)|0>,\\
\pi ( "det"(\p+L(z))) |0> ~ = ~ (\p+\l_n/z)(\p+\l_{n-1}/z)...(\p+\l_1/z)|0>
\eea
}

{\Rem  Actually one can prove that
the coefficients of the $1$-spin QCP belong
to the center of the universal enveloping algebra $U(\g)$ so it is true that:\\
$$\pi ( "det"(\p-L(z))) = (\p-\l_1/z)(\p-\l_2/z)...(\p-\l_n/z) Id$$
and the same  for  $\pi ( "det"(\p+L(z)))$. These formulas obviously provide
an explicit description for the Harish-Chandra map.}

{\Rem  This conjecture is  related to the
much more general results of E. Frenkel, E. Mukhin, V.Tarasov,
and A. Varchenko.}
\\~\\
By a straightforward calculation we obtain
{\Prop The conjecture is true for the $\mathfrak{gl}_2$ case.
}
\\~\\
Let us remark how the shifted action of the Weyl group,
well-known in Harish-Chandra homomorphism theory,
can be represented by different factorizations of QCP.
The images of Casimir elements under
the Harish-Chandra homomorphism are symmetric functions with respect
to the action of the Weyl group. The action
is given not by the naive formula $(\l_i)\to (\l_{\sigma(i)})$,
it is shifted. In the $\mathfrak{gl}_2$-case the
Weyl group is $S_2=<1,p>$ which acts by $p(\l_1,\l_2)=(\l_2-1, \l_1+1)$.
\\~\\
{\bf Observation} The same action of the permutation group $S_2$
arises considering different factorizations of the QCP:
\bea
(\p-\frac{\l_1}{z}) (\p-\frac{\l_2}{z})
=
(\p-\frac{\l_2-1}{z}) (\p-\frac{\l_1+1}{z})
\eea

\section{Appendices}
\subsection{Standard Lax operator for $\g[t]$}
Let us recall some definitions for the reader's convenience.

{\Def ~}
The standard Lax operator for $U(\g[t])/(t^N=0)$
($N$ can be finite or infinite)
and the standard Lax operator for $U(\g[t])/(t^N=0)$
with a constant term $K$ are given by the formulas:
\bea
\label{Lax-n}
L(z)=\sum_{i=0}^{N-1} \Phi_i z^{-(i+1)}, ~~~~~~~
L_K(z)=K+\sum_{i=0}^{N-1} \Phi_i z^{-(i+1)}.
\eea
Here $K$ is an arbitrary constant matrix,
$z$ - a formal parameter (not the same as $t$ !),
$$\Phi_i\in \Bigl( Mat_n\otimes U(\g[t])/(t^N=0) \Bigr) \cong  \Bigl( Mat_n[U(\g[t])/(t^N=0)] \Bigr)$$
are given by
\bea
\label{not1}
\Phi_i=\sum_{kl} E_{kl}\otimes e_{kl}t^i \Longleftrightarrow (\Phi_i)_{kl}=e_{kl}t^i
\eea
where
$ E_{kl}\in Mat_n$,  $e_{kl}t^i\in U(\g[t]).$ Both $E_{kl}$ and $e_{kl}$
are the matrices with the only $(k,l)$-th nontrivial matrix element equal
to $1.$ We consider them as elements of different
algebras: the first is an element of the associative algebra $Mat_n$, the second -
of the universal enveloping algebra $U(\g[t])$.
\\
It is well-known that the Gaudin Lax operator given by the
formula \ref{main-part-Lax-Gaud}, as well as $L(z), L_K(z)$ satisfy the same commutation relation:
{\Prop
\bea
\label{rel1}
[\one L(z) , \two L(u) ] = [ \frac{P}{z-u}, \one L(z)  + \two L(u)  ]
\eea
}

{\Rem ~} The Lax operator $L(z)=K+\Phi/z$ defines an integrable
system on coadjoint orbits of $\g$. This system
is sometimes called the Mishenko-Fomenko system, and the method
to obtain it - the "argument translation method".

\subsection{Conjugation to Miura form}
{\Prop Consider the differential operator:
 $$\p^n-\sum_{i=1}^n H_i(z)\p^{n-i}=(\p-\chi_n(z))...(\p-\chi_1(z))$$
then the Drinfeld-Sokolov connection
\bea
\p-L_{DS}=
\p-\left(
\begin{array}{cccccc}
0 & 1 & 0 & 0 &... &0 \\
0 & 0 & 1 & 0 &... &0 \\
... & ... & ... & ...&... &... \\
0 & 0 & 0 & ... & 0 &1 \\
H_{n} (z) & H_{n-1}(z) & H_{n-2}(z) & ... & H_2(z) & H_1(z)
\end{array}\right)
\eea
is gauge equivalent to the Miura connection
\bea
(\p - L_{M}(z))=\p -
\left(
\begin{array}{cccccc}
\chi_1(z) & 1 & 0 & 0 &... &0 \\
0 & \chi_2(z) & 1 & 0 &... &0 \\
... & ... & ... & ...&... &... \\
0 & 0 & 0 & ... & \chi_{n-1}(z)&1 \\
0 & 0 & 0 & ... & 0 &\chi_{n}(z)
\end{array}\right).
\eea
Moreover if the vector $\Psi(z)$ satisfies the equation
$(\p-L_M(z))\Psi(z)=0$, then its first component provides a solution
of the equation $(\p-\chi_n(z))...(\p-\chi_1(z)) \Psi_1(z)=0$.
}
\\~\\
{\bf Proof~} We proceed by showing that there exists such an element $B$
which transform the connection $\p-L_{DS}$ to the connection of the Miura
form $\p-L_M.$ Let $S$ be a solution of the equation
\bea
\p S =
\left(
\begin{array}{cccccc}
\chi_1 & 1 & 0 & 0 &... &0 \\
0 & \chi_2 & 1 & 0 &... &0 \\
... & ... & ... & ...&... &... \\
0 & 0 & 0 & ... & \chi_{n-1}&1 \\
0 & 0 & 0 & ... & 0 &\chi_{n}
\end{array}
\right)S\eea
(we omit the dependence on the parameter $z$ for simplicity). One has
\bea
\p S_1&=&\chi_1 S_1+ S_2\nn\\
\p^2 S_1&=& (\p \chi_1+\chi_1^2) S_1+(\chi_1+\chi_2) S_2 +S_3\nn\\
etc.&&\nn
\eea
Hence by a linear change of basis with the lower triangular matrix $B$ of
the form
\bea
B=\left(
\begin{array}{cccccc}
1 & 0 & 0 & 0 &... &0 \\
\chi_1 & 1 & 0 & 0 &... &0 \\
\p \chi_1+\chi_1^2 & \chi_1+\chi_2 & 1 & ...&... &... \\
... & ... & ... & ... & 1&0 \\
... & ... & ... & ... & ... &1
\end{array}
\right)
\eea
one transforms the connection of the Miura type $\p-L_M$ to some
connection of the DS type. The only thing to prove is that this connection
is exactly $\p-L_{DS}.$ To do this one has to realize that the condition
$(\p-L_{DS})\tilde{S}=0$ is equivalent to the condition $"det"(\p-L)\tilde{S}_1=0.$ Let us show
that $S_1$ solves the equation $"det"(\p-L){S}_1=0.$ Indeed,
\bea
S_2&=&(\p-\chi_1)S_1\nn\\
S_3&=&(\p-\chi_2)(\p-\chi_1)S_1\nn\\
...\nn\\
0&=&(\p-\chi_n)\ldots(\p-\chi_1)S_1\nn
\eea
$\square$

\end{document}